\documentclass[reprint,aps,prb,amsmath,amssymb]{revtex4-2}
\usepackage{graphicx}
\usepackage{hyperref,subfigure}
\usepackage{comment,array}
\usepackage[version=4]{mhchem}
\usepackage{physics}
\usepackage{dcolumn}
\usepackage{bm}
\usepackage{soul,xcolor}
\usepackage{ulem}
\setstcolor{red}
\usepackage{tabularray}

\usepackage{orcidlink}

\usepackage[flushleft]{threeparttable}
\usepackage{array,booktabs,makecell}

\usepackage{chemarrow}
\newcommand{\UD}[1]{{\color{magenta}{#1}}}

\newcommand{\FM}[1]{{\color{blue}{#1}}}

\newcommand\rst{\bgroup\markoverwith{\textcolor{red}{\rule[0.5ex]{2pt}{1.6pt}}}\ULon}

\newcommand\Ust{\bgroup\markoverwith{\textcolor{black}{\rule[0.5ex]{2pt}{1.6pt}}}\ULon}

\date{\today}
\begin{document}

\title{Pairing symmetry and superconductivity from long-range Coulomb interactions in the extended $t$-$t'$-$t''$-$J_z$ model for cuprates}
\author{U. A. Diaz-Reynoso\orcidlink{0000-0002-5172-7352}}
    \email[Correspondence email address: ]{udiazr@ens.cnyn.unam.mx}
    \affiliation{Departamento de Física, Centro de Nanociencias y Nanotecnología, Universidad Nacional Autónoma de México, Apartado Postal 14, 22800, Ensenada, Baja California, México}

\author{F. Mireles\,\orcidlink{0000-0003-0506-0793}}
    \email[Correspondence email address: ]{fmireles@ens.cnyn.unam.mx}
    \affiliation{Departamento de Física, Centro de Nanociencias y Nanotecnología, Universidad Nacional Autónoma de México, Apartado Postal 14, 22800, Ensenada, Baja California, México}

\date{\today}

\begin{abstract}
We investigate the pairing symmetry of the ground-state phase diagram of an extended two-dimensional  $t$-$t'$-$t''$-$J_z$ model, in which the first-, second-, and third-nearest neighbor electron hopping terms ($t,t'$ and $t''$) and an anisotropic Ising-like antiferromagnetic interaction $J_z$ are treated on the same framework. We find that the dominant pairing symmetry depends sensitively on the sign and magnitude of the hopping parameters $t'$ and $t''$, showing pure $p$-wave, pure $d$-wave, or coexisting pairing channels, highlighting the decisive role of these terms.   
We further explore the role of the long-range and short-range repulsive interactions in the formation of pairs and analyze the specific case of hole-doped cuprates. Our results indicate that repulsive interactions favors hole pair escaping of the stripe domains. Furthermore, pairing correlation calculations strongly suggest that repulsive Coulomb interactions drive reentrant superconducting behavior at experimentally observed underdoped regime ($\delta\approx0.07$ to $0.15$), as in Neodymium-based cuprates. 
\end{abstract}

\maketitle

\section{Introduction}
Strongly correlated materials, such as high-$T_c$ superconducting cuprates remain of current interest as their phase diagram is still not fully understood\cite{Phillips_2022,Cooper_2009}. The phase diagram feature intertwined  antiferromagnetic, pseudogap, strange metal, and superconducting phases, whose interplay continues to be the subject of active debate.~\cite{Keimer_2015,Armitage_2010,Lee_2006,Huang_2019,Tanaka_2026}. 
Simplified theoretical models
such as the Hubbard and $t$-$J$ models have succeeded in describing several aspects of quasi-two-dimensional cuprates, as competing inhomogeneous orders in the underdoped regime, including superconductivity~\cite{Emery_1987,Corboz_2011,Jiang_2022,Jiang_2021} and the appearance of unidirectional stripe ordering \cite{Footnote1}
%{Stripe ordering correspond to unidirectional modulations of charge and spin density that emerge in the underdoped regime, particularly near $\delta=1/8$, where holes segregate into one-dimensional domain walls separated by antiferromagnetically ordered regions.} 
-- corresponding to a self-induced antiphase domain walls of antiferromgnetic spins~\cite{Tranquada_1995,Chernyshev_2002,Kivelson_2003,Qin_2020,Corboz_2014,B.X.Zheng_2017,Xu_2024}.
Nevertheless,  most models fail to reproduce superconductivity over a wide doping range for hole-doped cuprates, particularly the  the well known  anomaly near the hole doping of $\delta \approx 1/8$. %\FMCom{[If  {\bf most models} fails, that statement implies that there are some that do not. Then the latter are the $good$ models. Paraphrasing is needed to better convey the idea- ]}
Although some variants that include longer-range hoppings, such as $t'$ and $t''$, successfully capture pairing on the electron-doped side~\cite{Jiang_2021,Jiang_2022}, the hole-doped regime systematically yields non-superconducting stripe ground states, suggesting that additional physical ingredients are essential~\cite{Qin_2020,Jiang_2024}. 

Experimental reports on Nd-doped La$_{2-x}$Sr$_x$CuO$_4$  (Nd-LSCO) indicate superconducting behavior across a wide range $0.055<\delta<0.20$, with a notable suppression at $\delta\approx 0.12$. This suppression is accompanied by the stabilization of static charge and spin stripe order, as observed via neutron scattering and muon spin rotation experiments~\cite{Matsuda_2002,Tranquada_1995,Niedermayer_1998}. At this critical doping, parallel-ordered stripes ~\cite{Ma_2022} emerge, suggesting their competition and coexistence with the superconducting phase ~\cite{Wen_2019,Arpaia_2019,Tranquada_2020}. %\FMCom{[Please check references marked with ? sign here]}\UDCom{[\bf{DONE}]}
This puzzling behavior poses a substantial challenge to theoretical models and numerical studies of high-$T_c$ superconductivity.

A common approach to include electron correlations in simplified lattice models is through the inclusion of an on-site Hubbard repulsion $U$.
However, \textit{ab initio} calculations ~\cite{Calzado_2001,Hirayama_2018}  and experimental  \cite{Chen_2021,Yamase_2026}  results points that the nearest-neighbor repulsion $V$ can be also significant in LSCO compounds, with typical ratios $|V/U| \approx 0.15$--$0.6$.
Including only short-range repulsive interactions has proven insufficient; at the estimated values of $V$, superconductivity is usually suppressed\cite{Farkasovsky_2026}. Earlier studies of long-range repulsive interactions were performed with the aim of avoiding phase separation ~\cite{Aligia_1995,Riera_1994,Troyer_1993}. Nevertheless, phase separation can be avoided without such long-range terms; stripe phases were observed as the ground state in powerful density-matrix renormalization group (DMRG) calculations on long cylinders ~\cite{White_1998,White_2000}. The stripe ordering found in these calculations were half-filled, consistent with the experiments of Tranquada et al.~\cite{Tranquada_1995}. This agreement strengthen confidence in the model and subsequently triggered the neglect of long-range repulsive interactions.

However, there is an issue with these striped solutions: they do not exhibit superconductivity in the expected range $\delta\le1/8$~\cite{Qin_2020}. The inclusion of second- and third-neighbor hopping amplitudes ($t'$ and $t''$) has allowed indeed a more comprehensive description of a wide range of phenomena in cuprates. For instance Jiang et al. ~\cite{Jiang_2021,Jiang_2022,Jiang_2024} mapped the ground- state phase diagram over a broad parameter space of $t'$ and $t''$. For electron-doped cuprates, they found $d_{x^2-y^2}$ superconductivity. For hole-doped cuprates, however, they observed only non-superconducting stripes, indicating that a key ingredient is still missing.

%Long-range repulsive interactions were explored in the 1990s with the goal of avoiding phase separation~\cite{Aligia_1995,Riera_1994,Troyer_1993}. Nevertheless, phase separation can be avoided without such long-range terms; stripes were observed using powerful density-matrix renormalization group (DMRG) calculations on long cylinders, which showed stripes as the ground state~\cite{White_1998,White_2000}. The stripes found in these calculations were half-filled, as observed in the experiments conducted by Tranquada et al.~\cite{Tranquada_1995}, consolidating confidence in the model and leading to the neglect of long-range repulsive interactions.

%However, there is a problem with these striped solutions: they do not exhibit superconductivity in the range $\delta\le1/8$~\cite{Qin_2020}. The inclusion of second- and third-neighbor hopping amplitudes ($t'$ and $t''$) has allowed a more comprehensive description of a wide range of phenomena in cuprates. Jiang et al.~\cite{Jiang_2021,Jiang_2022,Jiang_2024} mapped the ground-state phase diagram over a broad parameter space of $t'$ and $t''$. For electron-doped cuprates, they found $d_{x^2-y^2}$ superconductivity. For hole-doped cuprates, however, they observed only non-superconducting stripes, indicating that a key ingredient is still missing.

%%What interaction can preserve the stripe phase and display superconductivity at the same time?

The aim of this work is precisely to tackle this issue. To that end, we first perform a systematic investigation of the pairing symmetry in the ground-state phase diagram of the two-dimensional $t$-$t'$-$t''$-$J_z$ model, combining symmetry-based analysis with numerical calculations. We then extend the model by incorporating long-range repulsive interactions, and explore the mechanism by which superconducting pairs escape the non-superconducting stripes across a broad doping range, thus restoring the superconducting phase.

%In this work, we propose a minimal model that appears to resolve this problem: a $t$-$J_z$-$V$ model that incorporates second- and third-neighbor hopping, Ising-like magnetic interactions, and long-range repulsive interactions. Within this model, we show that superconducting pairs canO escape the non-superconducting stripes over a wide doping range, thereby giving rise to superconductivity. Additionally, we discuss the role of the parameters $t'$ and $t''$ in determining pair symmetry, using both numerical and analytical methods.

\section{Model Hamiltonian}

The minimal lattice Hamiltonian employed in this work is derived from the large-$U$ limit of the Hubbard model ($|U/t|\gg 1$), in which double occupancy is forbidden. The Hamiltonian comprises three additive terms: (i) a kinetic part accounting for first-, second-, and third-nearest neighbor hopping, $H_{t\text{-}t'\text{-}t''}$, (ii) a near neighbor Ising-type magnetic exchange term $H_{J_z}$, and (iii) a long-range Coulomb repulsive interaction, $H_V$.  It reads
\begin{align}
H = H_{t\text{-}t'\text{-}t''} + H_{J_z} + H_V.
\label{eq:hamiltonian}
\end{align}

The first two terms of the right conform the known ${t\text{-}t'\text{-}t''\text{-}J_z}$ model. 
\vspace{0.2cm}

%\subsection{Kinetic term}
 (i) {\it Kinetic contribution:} the kinetic part of the Hamiltonian (\ref{eq:hamiltonian})  is given by

\begin{align}
H_{t-t'-t''} &=
   \Bigg(-t\!\!\sum_{\langle \bm {i,j}\rangle\sigma} - t'\!\!\!\!\sum_{\langle\langle \bm i,\bm j\rangle\rangle\sigma}
     -  t''\!\!\!\!\!\!\sum_{\langle\langle\langle \bm i,\bm j\rangle\rangle\rangle\sigma}\Bigg) c_{\bm i\sigma}^\dagger c_{\bm j\sigma} + \text{h.c.}
\end{align}

%\begin{align}
%H_{t-t'-t''} &=    \left(-t\sum_{\langle i,j\rangle\sigma} - t'\sum_{\langle\langle \bm i,\bm j\rangle\rangle\sigma}  - t''\sum_{\langle\langle\langle \bm i,\bm j\rangle\rangle\rangle\sigma}\right) (c_{\bm i\sigma}^\dagger c_{\bm j\sigma} + \text{h.c.}) \end{align}
Here, $c_{\bm i\sigma}^\dagger$ and $c_{\bm i\sigma}$ are the creation and annihilation electron operators with spin $\sigma=\{\uparrow,\downarrow\}$ at the vector position  ${\bm i}=(x,y)$  being  $x$ and $y$ the particle coordinates in lattice units. The sums over $\langle \bm i,\bm j\rangle$, $\langle\langle \bm i,\bm j\rangle\rangle$, and $\langle\langle\langle \bm i,\bm j\rangle\rangle\rangle$ run over first, second, and third nearest neighbors bonds, with corresponding hopping amplitudes $t,t'$ and $t''$, respectively.  
%; as the $t$-$J$ model, double-occupied states are excluded.

The sign of $t'$ is related to the doping type: $t'>0$ for electron-doped and $t'<0$ for hole-doped cuprates. 
Thus, $t'$ is essential for capturing the electron-hole asymmetry. Notably, the estimated values in cuprates, $|t'/t|\approx 0.3$ and $|t''/t|\approx 0.2$~\cite{Calzado_2001,Leung_1997}, render these terms non-negligible. Although several studies have examined pairing symmetries in related models ~\cite{Kagan_1994,Fye_2004}, our analysis reveals how the sign and magnitude of $t'$ and $t''$ govern both the existence and the symmetry of pairs. Additionally, we demonstrate that certain symmetries forbid specific states, thereby modifying the pair characteristics.  

One might also consider the three-site hopping term. Nevertheless, Leung et al.~\cite{Leung_1997} showed that, although it provides minor refinements, it is not essential for reproducing Angle-Resolved Photoemission Spectroscopy (ARPES) data within the $t$-$t'$-$t''$-$J_z$ model. Given that this term acts on second- and third-neighbor sites, we believe its effects are effectively masked by $t'$ and $t''$ terms. For simplicity, we therefore omit it in the present work.

\vspace{0.2cm}

(ii) {\it Spin exchange interaction:} 
 Within the large-$U$ limit, the Hubbard model reduces via canonical transformation to the $t$-$J$ model, with superexchange $J=4t^2/|U|$. To reduce computational cost, we further adopt the Ising approximation ($J_z S_{\bm i}^z S_{\bm j}^z$) instead of the full vector exchange ($J \mathbf{S}_{\bm i} \cdot \mathbf{S}_{\bm j}$ ), yielding to an Ising-like magnetic interaction.  The nearest neighbor spin exchange Hamiltonian is thus  

%The \FM{spin} interaction used in this work is based on the $t$-$J_z$ model. 
%Within the large U limit of the Hubbard model can be simplified via a canonical transformation to the $t$-$J$ model. Here, the spin interactions are governed by the superexchange parameter $J=4t^2/|U|$. A further simplification can be achieved by using the $t$-$J_z$ model instead of the full $t$-$J$ model, replacing the vector spin exchange $J \mathbf{S}_{\bm i} \cdot \mathbf{S}_{\bm j}$ with its Ising approximation $J S_{\bm i}^z S_{\bm j}^z$, which is computationally less demanding.
%\FM{Thus the} Ising-like magnetic interaction \FM{is described by a nearest neighbor spin exchange} given by
\begin{align}
    H_{J_z} = J\sum_{\langle \bm i,\bm j\rangle}\left(S_{\bm i}^z S_{\bm j}^z - \frac{n_{\bm i} n_{\bm j}}{4}\right),
\end{align}
where the occupation operator $n_{\bm i} = n_{\bm i\uparrow} + n_{\bm i\downarrow}$, $n_{\bm i\sigma}=c_{\bm i\sigma}^\dagger c_{\bm i\sigma}$, and $S_{\bm i}^z=\frac 1 2 (n_{\bm i\uparrow}-n_{\bm i\downarrow})$ is the $z$-component of the quantum spin operator $\mathbf{S}_{\bm i}$. The $t$-$J_z$ model retains most of the key physical phenomena of the $t$-$J$ model, particularly pairing and stripe correlations~\cite{Riera_1993,Riera_2001}. In some cases, both models yield qualitatively and quantitatively similar results upon scaling $J$ by a factor~\cite{Ma_ka_2012}. Typical values of $J$ for cuprates are around $J\sim 0.3$--$0.5$~\cite{Sheshadri_2023,Jiang_2022}. We set $J=0.4$ in this work.

\vspace{0.2cm}
%\subsection{Repulsion term}
(iii) {\it Repulsive long-range interactions:}
Screening effects in metals and complex oxides are commonly  modeled through Yukawa-type potentials, $V(r)\propto e^{-r/\lambda}/r$, where $\lambda$ denotes the screening length. However, this result is rooted in the Thomas–Fermi approximation, whose underlying assumptions are free-particle behavior and three-dimensional isotropy. In contrast, La\(_{2-x}\)Sr\(_x\)CuO\(_4\) (LSCO) is a prototypical two-dimensional bad metal, characterized by an in-plane resistivity approximately 500 times that of copper, a highly anisotropic orthorhombic structure (\(c/a = 2.44\)), and an out-of-plane to in-plane resistivity ratio \(\rho_c/\rho_{ab} > 10^3\) ~\cite{Ando_1995}. 
Consequently, the weak interlayer coupling and large interplane spacing act to significantly suppress screening. Moreover, recent studies show that long-range interactions persist even in two-dimensional metallic systems~\cite{Ramezani_2024}, and a theoretical analysis of stripes under repulsive interactions concluded that long-range stripe order can arise from a generalized dipolar interaction \(1/r^{\alpha}\) only for \(\alpha < 2\)~\cite{Mendoza_Coto_2015}. These observations motivate us to adopt a long-range Coulomb repulsive potential of the form,

\begin{align}
    H_V = \sum_{\bm i \neq \bm j} \frac{V}{2|{\bm i}-{\bm j}|} n_{\bm i} n_{\bm j}.
\end{align}
The distance $|\bm r|=|{\bm i}-{\bm j}|$ (in lattice units) used in this work is that of the Clifford torus \cite{Vlachos_2005}. The Clifford torus is a two-dimensional flat manifold that can be regarded as the product of two circles embedded in a four-dimensional space. The induced metric is a smooth function of its coordinates. This coordinate mapping has been used previously in $t$-$J$ models with long-range interactions~\cite{Kuramoto_1991,Kuramoto_1993} and to efficiently calculate the energy of periodic and infinite charge distributions as well as the Madelung constant~\cite{Tavernier_2021,Zhao_2026}.
The distance on the Clifford torus is given explicitly by
\begin{align}
|\bm r| = |(x, y)| = \sqrt{x'^2 + y'^2},
\end{align}
\FM{} with $x',y'$ defined as
\begin{align}
    x'(x) & = \frac{L}{\pi} \sin(\pi |x| / L), \\
    y'(y) & = \frac{M}{\pi} \sin(\pi |y| / M),
\end{align}
where $L$ and $M$ are the length and width of the system (per lattice width). 
Hence $N = L \times M$ as the total number of lattice sites.
The value of $V$ can be inferred from \textit{ab initio} calculations~\cite{Calzado_2001}. From the difference between first- and second-nearest  neighbor potentials, $V_1-V_{\sqrt{2}}=1.44$ (in units of $|t|$), we obtain $V\approx 4.9$, while from the first- and third-neighbor potentials, $V_1-V_2=2.65$, we get $V\approx 5.3$.

\section{Numerical Method}
 \label{method} 
The ground-state properties of the extended Hamiltonian (\ref{eq:hamiltonian}) are obtained via a truncated Lanczos method\cite{Dagotto_1998,Chiappa_2001}. This approach combines a variational principle with iterative Hilbert-space reduction and exact diagonalization through the standard Lanczos algorithm applied in successive subspaces. 
The geometry of the system consist of a rectangular lattice comprising $N = L \times M$ sites, subject to fully periodic boundary conditions. 
The procedure starts from a trial state $\ket{\Psi_{\bm k}}$ in an $n$-dimensional Hilbert space and iteratively constructs eigenstates from a local basis of the form
\begin{align}
  \ket{\Psi_{\bm k}} = \sum_{\bm R} a_{\bm R} \ket{\bm k;\bm R} =  \sum_{\bm R} a_{\bm R}  \frac{ \hat{P}_{\bm k} }{\sqrt{N_{\bm k,\bm R}}} \ket{\bm R},
\end{align}
where $a_{\bm R}\in\mathbb C$, and \( \ket{\bm R} = c^\dagger_{\bm i_1,\sigma_1} c^\dagger_{\bm i_2,\sigma_2} \cdots \ket{0} \) describes a configuration state, where $c^\dagger_{\bm i_{\nu},\sigma_\nu}$ is the creation operator of a Wannier state of a $\nu$-th electron at the vector position $\bm i_{\nu}$ with spin $\sigma_{\nu}=\{\uparrow,\downarrow\}$. 
%We also can denote a configuration state as an array of $\uparrow,\downarrow$ and $0$ (unoccupied site) symbols, e.g. $\ket{\bm R}=\ket{\begin{matrix}\uparrow & 0&\downarrow&\uparrow\end{matrix}}$}. 
Here, $\hat P_{\bm k}$ denotes the projector operator defined by
 \begin{align}
    \hat P_{\bm k}=\frac 1 N \sum_{\bm r} e^{i \bm k\cdot \bm r } \hat T_ {\bm r},
\end{align}
\noindent such that $N_{\bm k,\bm R}=\bra{\bm R} \hat{P}_{\bm k} \ket{\bm R} \neq 0$, and  the operator $\hat T_{\bm r}$ translates a creation or annihilation operator from position $\bm i$ to $\bm i+\bm r$,  
\begin{align}
    \hat T_{\bm r} c^\dagger_{\bm i}\hat T^\dagger_{\bm r}=c^\dagger_{\bm i+\bm r}.
\end{align}
%\UD{The states $\ket{\bm k;\bm R}$ can be thought as a plane wave made of all translations of $\ket{\bm R}$ (modulated by $e^{i\bm k\cdot \bm r}$), e.g. for $\bm k=0, \ket{\bm k;\bm R}$ is simply the superposition of all translations of $\ket{\bm R}$ with the same amplitude.}
%%%\UD{The configuration state $\ket{\bm R}$ that has $|a_{\bm R}|\ge |a_{\bm R'}|$ for any other $\ket{\bm R'}$, is called the most probable configuration.}

At each iteration, the Hilbert space is enlarged by acting the Hamiltonian $H$ to the current state $\ket{\Psi_{\bm k}}$; which is  diagonalized within this expanded subspace, and then the space is truncated to retain only the $\gamma n$ states with the largest amplitudes ($\gamma>1$). This cycle is repeated until convergence. When combined with a block Lanczos extension, the method is used to obtain the energy of a few low lying states. The trial wavefunction is typically initialized as a single configuration $\ket{\bm R}$—for example, the N\'eel state with a few holes, or suitable configuration states candidates to the ground state. 

%then enlarging the Hilbert space by applying the Hamiltonian $H\ket{\Psi}$, diagonalizing $H$ restricted to this enlarged space, truncating the Hilbert space to the $\gamma n$ states with the largest amplitudes ($\gamma>1$), and repeating. This method can be used in combination with block Lanczos to obtain the energy of a few states.The trial wavefunction can contain only one state $\ket{\bm R}$ (usually the N\'eel state with a few holes). 

We found that for the systems analyzed, states with zero total momentum ($\bm k=0$), have the lowest energy. All systems analyzed have total spin $z$-component $S^z_{\text{total}}=0$ (equal number of $\uparrow$ and $\downarrow$ electrons).
The total energy of a state $\ket{\Psi_0}$ is given by $E=\bra{\Psi_0} H\ket{\Psi_0}$. The magnetic energy is $E_{J_z}=\bra{\Psi_0} H_{J_z}\ket{\Psi_0}$ and the Coulomb energy is $E_{V}=\bra{\Psi_0} H_{V}\ket{\Psi_0}$.
At $V=0$ and no holes, the ground state within this model for a lattice of size $L\times M$ with both $L,M$ even is the N\'eel state with energy $E_0=-LMJ$. For a lattice with $L$ odd and $M$ even, the ground state is similar to a N\'eel state but features a defect domain wall of length $M$, giving $E_0=-LMJ+MJ/2$.
The standard practice is to measure the energy per hole rather than the total energy. 
%\FMCom{[Note that for consistency of the notation of $\ket{\Psi_{\bm k}}$, I have added the subscript $0$ to the states, that is, $\ket{\Psi}\rightarrow{\ket{\Psi_0}}$.]} 
The energy per hole is thus defined as

\begin{align}
  {\cal E}=\begin{cases}
    \left(E-E_0\right)/n_h \quad &\text{for $L,M$ even}\\
    \left(E-E_0+MJ/2\right)/n_h \quad &\text{for $L$ odd, $M$ even}
  \end{cases}
\end{align}

%\begin{align}  e=\begin{cases}  \left(\bra{\Psi_0}H\ket{\Psi_0}-E_0\right)/n_h \quad &\text{for $L,M$ even}\\   \left(\bra{\Psi_0}H\ket{\Psi_0}-E_0+MJ/2\right)/n_h \quad &\text{for $M$ even,$L$ odd}  \end{cases} \end{align}

\noindent where $n_h$ is the number of holes and $E_0$ is the ground state energy without holes. We define the electric charge energy per hole as ${\cal E}_V=E_V/n_h$ and the $t$-$J_z$ energy per hole as ${\cal E}_{t-J_z}={\cal E}-{\cal E}_V$, which gives the energy (per hole) of the system in the absence of Coulomb electron repulsion.

%\FMCom{[$h_ {\bm r}$ is the hole nuber operator at site ${\bm r}$]}

%\UD{ Also, we analyze the charge-charge correlation function, more precisely, the hole-hole correlation function. For a system with periodic boundaries we have: \begin{align} \langle n_{\bm i }n_{\bm j} \rangle-\langle n\rangle^2=\langle h_{\bm i}h_{\bm j}\rangle -\langle h\rangle^2=\langle h_{\bm r}h_{\bm 0}\rangle-\langle h\rangle ^2 \end{align} where $\langle n\rangle (\langle h\rangle)$ is the average number of electrons (holes) per-site, $\bm r= \bm i-\bm j$ and the charge-charge and hole-hole correlation function give us the same information. We have opted to report $N\langle h_{\bm r} h_{\bm 0}\rangle$ instead, we have that: $\sum _{\bm r}N\langle h_{\bm r}h_{\bm 0}\rangle=\langle h^2\rangle$}\FMCom{[{\bf We need to discuss about this.}]}

%\FM{The hole-hole correlation function is evaluated through the standard formula $\sum _{\bm r}N\langle h_{\bm r}h_{\bm 0}\rangle=\langle h^2\rangle$, where $h$ is the number operator for holes.}

\section{Results and discussion}
%In this section we present the results obtained. 
We begin by analyzing the pair symmetry of two-hole states using theory of invariants and the aid of numerical computations. Then, we review the charge stripe phase and calculate some of its properties within the ${t\text{-}t'\text{-}t''\text{-}J_z}$ model. Subsequently, we study the model subject to long-range repulsive interactions; and finally, the quantum pair-pair correlations of the full model is investigated and discuss on the emergence of $d$- and $p$- wave superconductivity triggered by the repulsive interactions.
%\vspace{0.5cm}

%\subsection{Pair symmetr}}

\subsection{Symmetries in a square lattice with two holes}
Let us consider the presence of two holes on a square lattice of size $M\times M$ 
%\FM{($M \in \mathbb{N}$)}\st{($M$ even)}
, in the absence of long-range electron interactions, and periodic boundary conditions.
%\FMCom{[Tanslations ?...]}.
A state $\ket{\bm k;\bm R}$ is invariant under $g \in G$, where $G$ is a symmetry group that leaves the Hamiltonian invariant, if it satisfies
\begin{align}
    \hat{D}_g \ket{\bm k;\bm R} = \lambda_g \ket{\bm k;\bm R},
\end{align}
where $\hat{D}_g$ is the unitary representation of $g$ and has eigenvalues $\lambda_g \neq 0$. Then $\ket{\bm k;\bm R}$ is also eigenstate of every $g' \in \langle g \rangle$, the cyclic subgroup generated by $g$. We call such states \textit{high-symmetry} states. 
For a lattice point group and a state with zero total momentum, $\ket{\bm k=\bm 0;\bm R}$; the value of $\lambda_g$ can only be $\pm 1$. More precisely, $\lambda_g = (-1)^{p_g}$, where $p_g$ is the number of electron transpositions (electron pair exchanged) allowed by the symmetry operation $g$. The point symmetry of the square lattice corresponds to the $D_4$ point group, generated by a $90^\circ$ clockwise rotation $r$ and a reflection $s$ about the bottom-left to top-right diagonal. The group has a total of five conjugacy classes, represented by the subsets 
$\{e\}$, $\{r,r^3\}$, $\{r^2\}$, $\{s,sr^2\}$, and $\{sr,sr^3\}$, (see Table I). 

For a $M\times M$ lattice containing two holes, the hole configuration may be invariant under a proper subgroup ${\cal H} \subset D_4$ of the full spatial point group, depending on the positions of the holes and if $M$ is even or odd. Three illustrative examples of such symmetries for the case of $M\ge 4$ with $M$ even are:
\vspace{0.2cm}

\noindent   ({\it a}) Invariance under reflections about the horizontal and vertical axes (denoted, for example, by $sr^3$ and $sr$, respectively, if $s$ is the diagonal reflection), and under a $180^\circ$ rotation $r^2$: $(x,y) \to (-x,-y)$ (Fig.~1(a)). The $180^\circ$ rotation leaves four electrons invariant at the coordinate indices $(0,0), (M/2,0), (0,M/2), (M/2,M/2)$. With two holes in the system (which are not exchanged under this operation), the number of exchanged electron pairs is (see Appendix A)
$$p_{r^2}=\frac{(M^2-6)}{2},$$
leading to $\lambda_{r^2}=-1$ for $M\ge 4$. This implies that $\lambda_{r^3}=-\lambda_r,\lambda_{s}=-\lambda_{sr^2}$ and $\lambda_{sr}=-\lambda_{sr^3}$.  Entailing that the operations in two-elements equivalence classes have opposite sign.

Under the vertical reflection $(x,y) \to (-x,y)$, and for even \(M\), the operation leaves two columns invariant: $x=0$ and $x=M/2$ (owing to the periodic boundary condition $-x \equiv M-x \pmod M$. Assuming the two holes are placed in these fixed columns, they are not exchanged. The remaining $M-2$ columns form $(M-2)/2$ exchanged column pairs. Since each column contains \(M\) sites, the total number of exchanged electron pairs is (see Appendix A)
$$
p_{sr^3} = M \times \frac{M-2}{2} = \frac{M}{2}(M-2).
$$
This number is always even for even $M\ge 4$, hence the eigenvalue of the reflection operator is \(\lambda_{sr^3} = (-1)^{p_{sr^3}} = 1\).

({\it b}) Invariance under reflection about the vertical axis ($sr$); here the number of exchanged pairs is again $p_{sr}= M(M-2)/2$, giving $\lambda_{sr}=1$, note that in this case the first column 0 and $M/2$ are invariant, (Fig. ~1(b)).

({\it c}) Invariance under reflection about the bottom-left to top-right diagonal ($s$). In this special case, only the sites lying on the diagonal are invariant; the two holes are pinned to their sites and therefore are not exchanged
This yields $p_{s}=(M^2-M-2)/2$ exchanged electron pairs. The corresponding eigenvalue $\lambda_s$ takes the value  $-1$ when $M/2$ is even, and $+1$ when $M/2$ is odd, (Fig.~1(c)). Clearly the case $\lambda_s=-1$ implies $\lambda_{sr}=-\lambda_r,\lambda_{sr^2}=-\lambda_{r^2},$ and $\lambda_{sr^3}=-\lambda_{r^3}$.

%\noindent \FM{({\it i}) {\it Case $M$ even.}}
\begin{figure}[h]
  \includegraphics[page=4]{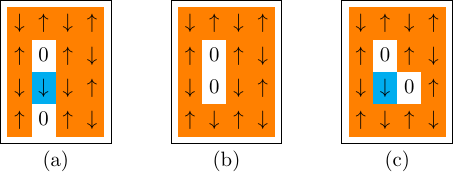}
  \caption{Three high-symmetry two-hole configurations in a $4\times4$ antiferromagnetic lattice. 
  The staggered magnetization $S^z_{i,j}  = (-1)^{i+j} S^z(i,j)$ is indicated with a background color in orange/blue for positive/negative values. Here, $S^z(i,j)=\pm 1/2$ for spin $\uparrow,\downarrow$ (respectively) is the on-site magnetization at site $(i,j)$, with $i=j=-1$ at bottom-left corner.
  %Here, $S^z(i,j)$ is the on-site magnetization for the site with $(i,j)$ indices and $i=j=-1$ at bottom-left corner.
 Empty sites (holes) are blank and marked with $0$. 
Using the $D_4$ generators defined in the text: (a) is invariant under $sr^3$ (horizontal) and $r^2$ ($180^\circ$ rotation), implying $sr$ (vertical); (b) is invariant under $sr$; (c) is invariant under $s$.} 
\label{fig:states1}
\end{figure}

To determine the symmetry of the ground state, we note that the Hamiltonian $H$  (Eq. \ref{eq:hamiltonian}) defined in a square lattice  can be block-diagonalized into subspaces corresponding to the irrep of the spatial (non-magnetic)  $D_4$ symmetry group. The projection onto a given irrep $\Gamma$ is obtained using the projection operator, 
\begin{align}
    \hat{P}_\Gamma = \frac{d_\Gamma}{|G|}\sum_g \chi_g^{\Gamma} \hat{D}_g,
\end{align}
where $d_\Gamma$ is the dimension of the irrep, $|G|$ is the order of the group, $\chi^\Gamma_g$ is the character of $g$ for the irrep $\Gamma$, and $\hat{D}_g$ is the representation of $g$. The character table for the $D_4$ symmetry group is shown in Table \ref{tab:irrep}.

%\FMCom{[ I believe that the standard notation in solid state physics for the irrep are the  Mulliken symbols $A_1$, $B_1$, E, ... instead of using $\chi_1$, $\chi_2$,...  ]}

%\FMCom{[Also, it seems that in general the character table for class $r$ is also for $r^3$, so you can write also \{ $r,r^3$\}  instead for just $r$, similarly for $s\rightarrow{\{s,sr^2\}}$ and ${\{sr\}}\rightarrow {\{s,sr^3\}}$. {\bf Please check this.}]}

\begin{table}[h!]
\centering
\[
\begin{array}{|c|ccccc|c|}
\hline 
\text{Irrep} & \{e\} & \{r, r^3\} & \{r^2\} & \{s, sr^2\} & \{sr, sr^3\} &\text{Basis functions} \\ \hline\hline
\chi_1(A_1) & 1 & 1 & 1 & 1 & 1 & x^2+y^2\\
\chi_2(A_2) & 1 & 1 & 1 & \!\!\!\!-1 & \!\!\!\!-1 & \\
\chi_3(B_1) & 1 & \!\!\!\!-1 & 1 & 1 &\!\!\!\! -1 & xy\\
\chi_4(B_2) & 1 &\!\!\!\! -1 & 1 &\!\!\!\! -1 & 1 & x^2-y^2\\
\chi_5(E)  & 2 & 0 & \!\!\!\!-2 & 0 & 0 & \{x, y\}
\\
\hline\hline
\end{array}
\]
\caption{Character table for the point symmetry group $D_4$. Columns denote conjugacy classes (with representative elements in braces). The rightmost column lists the basis functions transforming according to each irreducible representation.\cite{Dresselhaus2008}
}  
\label{tab:irrep} 
\end{table}

\vspace{0.5cm}
\subsection{Results for the $t$-$J_z$ model with two holes}

Using the numerical methodology described in Sec.~\ref{method}, we calculate the ground state of the $t$-$J_z$ model (with $t' = t'' = 0$, $\bm k = 0$, and $J_z = 0.4$) on an $M \times M$ two-dimensional lattice. We find that the ground state is two-fold degenerate for $M = 4, 6, 8, 10, 12$. The configuration shown in Fig.~\ref{fig:states1}(a), which we denote as 
$|\phi_{\scriptscriptstyle{(a)}}\rangle$, has a non-zero amplitude. Indeed, it is the most probable configuration for one of the two degenerate ground states.Note that the case $M=6$ has been already studied by \cite{Riera_1993}, though they do not discuss on the symmetries of the ground state.
%\UDCom{[Here, we may say that the found ground state for the $6\times 6$ lattice is exactly the same found by Riera et al. 1993 (the same energy up to 4 significant figures).]}

%This state can be regarded as the extension/continuation of (a) with the same antiferromagnetic background (colored in orange) to fit $M\times M$ and also has the same symmetries. 

The state $|\phi_{\scriptscriptstyle{(a)}}\rangle$ is characterized by the fact that  $\hat P_\Gamma
|\phi_{\scriptscriptstyle{(a)}}\rangle=0$ for every irreducible representation $\Gamma$ except for $\Gamma=\chi_5$. Since at least one of the two degenerate ground states has a non-zero projection onto $|\phi_{\scriptscriptstyle{(a)}}\rangle$, we conclude that the ground state belongs to the $\chi_5$ (i.e., $p$-wave) representation. Furthermore, the horizontal reflection preserves this state, while the vertical reflection changes its sign: $\hat D_{sr}|\phi_{\scriptscriptstyle{(a)}}\rangle = |\phi_{\scriptscriptstyle{(a)}}\rangle$ and $\hat D_{sr^3}|\phi_{\scriptscriptstyle{(a)}}\rangle= -|\phi_{\scriptscriptstyle{(a)}}\rangle$. This identifies 
$|\phi_{\scriptscriptstyle{(a)}}\rangle$ as having $p_y$-like character, with its $90^\circ$-rotated partner exhibiting $p_x$-like character. Hence, from the numerical solution of the full degenerate ground state we can construct two orthogonal eigenstates: one with zero projection onto $|\phi_{\scriptscriptstyle{(a)}}\rangle$ -the domain of Fig.1(a)-, 
denoted $\ket{p_x}$, and a second one corresponding to its orthogonal partner, denoted $\ket{p_y}$.

%\FM{Now, from Table I we can infer that due the opposite sign in the two-element equivalence classes and to the change of sign of the character between $\{e\}$ and $\{r\}$, the only irrep that allows a non-zero amplitude for a state with configuration as in Fig.1(a) is the $\chi_5 (E)$ representation - which have $\{x,y\}$ as it associated basis functions. Consequently its state configuration has $p_y$-like symmetry, whiles its $90^\circ$ rotated state has $p_x$-like symmetry. Hence, from the numerical solution of the full degenerate ground state we can construct two orthogonal eigenstates: one with zero projection onto configuration Fig.1(a), denoted $\ket{p_x}$ and a second one corresponding to its orthogonal partner $\ket{p_y}$.}  \UDCom{}

On the other hand, we find that the first excited state is non-degenerate and therefore transforms according to a one-dimensional irreducible representation. Numerically, this state exhibits non-zero amplitudes for the configurations shown in Figs.~1(b) and 1(c), both are invariant under specific symmetry operations of $D_{4}$, which determines the irrep to which this state belongs. 

%\UDCom{[Fig1b and Fig2b do not share the same symmetries. Im not sure what you wanted to say.]} \FMCom{[I believe that there is no space for misinterpretation here. My statement is clear, that both Figs.~1(b) and 1(c) have the same symmetries. That is what clearly is saying, nothing else. Now, please note that I was trying to interpret and make sense of what you wrote here before, and of what you told me about these cases. If such statement is not actually accurate after all, then please establish clearly and unambiguously what would be then the expected symmetries for Figs.~1(b) and 1(c), if they have some. Note that the main idea/reason of leaving this paragraph is to say something (if we can)  on the expected symmetries of the excited states, since we have already talked on the symmetries for the numerical ground state.]}

For a one-dimensional irrep a symmetry operation $\hat D_g$ acts on a state $|\Psi\rangle$ belonging to the irrep $\Gamma$ simply by the multiplication with the character $\chi_g^\Gamma$ \cite{Dresselhaus2008}:
\begin{align}
\hat D_g\ket{\Psi}=\chi_g^\Gamma\ket{\Psi}.
\end{align}

Now consider the configuration Fig.1(b), embedded on a $M\times M$ the antiferromagnetic background with $M$ even $\ge 4$. Denoting this configuration as state $|\phi_{\scriptscriptstyle{(b)}}\rangle$, and the first excited state as $\ket{\Psi}$, the matrix element of the reflection operator $\hat D_{sr}$ is given by  
\begin{align}
\bra{\phi_{(b)}} \hat D_{sr} \ket{\Psi}
= \chi_{sr}^\Gamma \braket{\phi_{(b)}}{\Psi}.
\end{align}

If $\ket{\phi_{(b)}}$ is an eigenstate of $\hat D_{sr}$ with eigenvalue $\lambda_{sr}$, i.e., $\hat D_{sr}\ket{\phi_{(b)}} = \lambda_{sr}\ket{\phi_{(b)}}$, then a non-zero projection $\braket{\phi_{(b)}}{\Psi}$ requires $\lambda_{sr} = \chi_{sr}^\Gamma$. Consequently, the only one-dimensional irreps allowing a non-zero amplitude for $\ket{\phi_{(b)}}$ are precisely those satisfying this condition, namely $\chi_1$ and $\chi_4$.

%Then, for the extension of (b) with the same antiferromagnetic domain in the $M\times M$ lattice, denoted $|\psi_{(b)}\rangle$, and this excited state $\Psi$, we have 

Similarly, consider configuration (c) of Fig.1 embedded in the same $M\times M$ antiferromagnetic lattice. Denoting this configuration state as $\ket{\phi_{(c)}}$, and the first excited state as $\ket{\Psi}$, the matrix element of the diagonal reflection $\hat D_s$ is given by
\begin{align}
\bra{\phi_{(c)}} \hat D_s \ket{\Psi}=\chi_{s}^\Gamma \braket{\phi_{(c)}}{\Psi}.
\end{align}

If $\ket{\phi_{(c)}}$ is an eigenstate of $\hat D_s$ with eigenvalue $\lambda_s$, i.e., $\hat D_s \ket{\phi_{(c)}} = \lambda_s \ket{\phi_{(c)}}$, then a non-zero projection $\braket{\phi_{(c)}}{\Psi}$ requires $\lambda_s = \chi_s^\Gamma$. As established earlier, $\lambda_s = -1$ when $M/2$ is even, and $\lambda_s = +1$ when $M/2$ is odd. Consequently, the state belongs to the $\chi_4$ irrep for $M/2$ even (yielding $d_{x^2-y^2}$ symmetry) and to the $\chi_1$ irrep for $M/2$ odd (yielding $s$-wave symmetry). We reserve the notation $\ket{d_{x^2-y^2}}$ specifically for the even-$M/2$ case.

In Fig.\ref{fig:pares} we show the numerical computations of the hole-hole correlation function $C_h({\bm r })=N\langle \hat h_{\bm r}\hat h_{\bm 0} \rangle$ of the  $t$-$J_z$ model ($t=1$ and $J_z=0.4$) in a $8\times8$ lattice with two holes for the ground ($\ket{p_x}$)
%\UDCom{[I would delete -like in both cases]} \FMCom{[Do not do it without a reasoned argumentation -- intended for me. Not here in the paper]}
%\UDCom{[In the Fig.2 we are showing hole-hole correlation function of  $\ket{p_x}$ and $\ket{d_{x^2-y^2}}$ states, I'm not sure why you use "-like" (maybe from a strictly mathematical point of view $\ket{p_x}$ is $\ket{p_x}$-like]} 
and first excited state ($\ket{d_{x^2-y^2}}$). Here $\hat h_ {\bm r}$ is the hole number operator at site ${\bm r}$ and $N$ is the total number of lattice sites. Note that the $p$-wave state yields a dipolar correlation pattern, while a  $d_{x^2-y^2}$ state yields a quadrupolar one. 
The observed symmetry of $C_h({\bm r })$ is not a numerical artifact but a direct consequence of the transformation properties of the state. Since the density operator is a scalar under the point group, the correlation function inherits the same irreducible representation as the state from which it is computed.   

%\UD{In the Fig.\ref{fig:pares} we show the charge-charge correlation function $\langle h_{\bm r}h_{\bm 0} \rangle$ for the $\ket{p_x}$ and $\ket{d_{x^2-y^2}}$ paired states.}

\UD{\begin{figure}[h]
\includegraphics[width=0.45\textwidth,page=34]{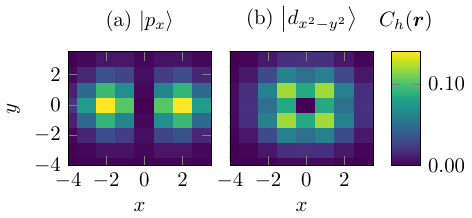}
\caption{Hole-hole correlation function $C_h(\bm r)$ for (a) the $\ket{p_x}$ component of the degenerate ground state, and (b) the first excited state $\ket{d_{x^2-y^2}}$. In each case, $C_h(\bm r)$ exhibits the same spatial symmetry as the corresponding state. }
\label{fig:pares}
\end{figure}
 }
 
\begin{figure}[h]
  \includegraphics[page=5]{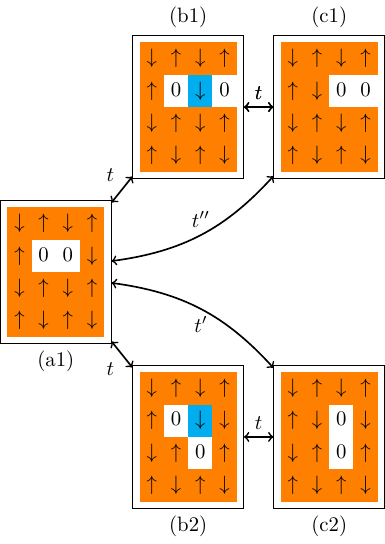}
  \caption{Schematic depiction of the hopping sequences for hole-pair propagation. The $\ket{p_x}$ state propagates by $(1,0)$ via the $t$-mediated sequence (a1)-(b1)-(c1), while the $\ket{d_{x^2-y^2}}$ state propagates by $(1/2,1/2)$ via the sequence (a1)-(b2)-(c2). The inclusion of non-zero $t'$ and $t''$ creates three-site loops that can lead to geometric frustration, inhibiting certain pairing symmetries.  %$t'<0$ hinders the movement of $\ket{d_{x^2-y^2}}$ and $t''<0$ hinders the movement of $\ket{p_x}$ and $\ket{p_y}$. %%Given that the cycle (a)-(c1)-(c2) exhanges a pair of electrons, this cycle can be represented by the hopping sequence $t'',-t',t'$.
  }
  \label{fig:propagation}
\end{figure}

%%\UD{With $t',t''\neq 0$ other channels of movement are available to the holes. Note that the movement mediated by $t'$ and $t''$ hopping, does not change the sublattice where the hole is (the site parity $(-1)^i+j$ is conserved) and then they won't be attached by the string of wrong antiferromagnetic bonds that are created when they move apart. Also, depending on the symmetry of the pair, the hopping $t'$ and $t''$ will impact of these states making them more or less thermodynamically favorable. For example, at $J=0.4,t=1,t'=t''=0$ and $V=0$, the difference of energy per-hole for the $\ket{d_{x^2-y^2}}$ and $\ket{p_x}$ is about $0.045t$. !!!!With $t'=-0.2$ and $t''=0$ the $\ket{d_{x^2-y^2}}$ is no longer the first excited nor the ground state and with $t'=0$ and $t''=0.05$ both pairs have practically the same energy. }

\subsection{Effect of $t'$ and $t''$ in the $t$-$J_z$ model}
 
We now explore the effect of $t'$ and $t''$ on the ground-state symmetry and on hole-pair mobility, and subsequently examine the corresponding pairing-symmetry phase diagram. When finite values of $t'$ and $t''$ are included, additional hopping channels become available to the holes. Depending on the pair symmetry, these hoppings modify the energies and wavefunctions of the states, rendering them energetically more or less favorable.

\begin{figure}
    \centering
    \includegraphics[width=0.48\linewidth,page=35]{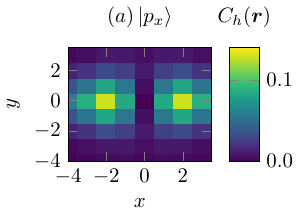}
    \includegraphics[width=0.48\linewidth,page=36]{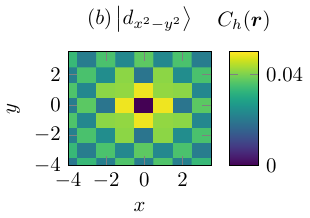}
    \caption{Hole-hole correlation function $C_h(\bm r)$ evaluated with the $t'$ and $t''$ hopping terms. (a) Paired $\ket{p_x}$ state with $t'=-0.3$ and $t''=0.2$; (b) $\ket{d_{x^2-y^2}}$ state with $t'=0.3$ and $t''=-0.2$. The rest fo the parameters are as in Fig.~\ref{fig:pares}. %Compared with Fig.~\ref{fig:pares}(b), the $\ket{d_{x^2-y^2}}$ state is significantly modified by the additional hopping, resulting in a less localized pair. In contrast, the $\ket{p_x}$ state remains virtually unchanged by $t'$ and $t''$.
    }
    \label{fig:waves-modified}
\end{figure}

In Fig.~\ref{fig:waves-modified} we show the hole-hole correlation function $C_h(\bm r)$  for, ~\ref{fig:waves-modified}(a) the paired $\ket{p_x}$ state with $t'=-0.3$ and $t''=0.2$ and for $\ket{d_{x^2-y^2}}$ state with $t'=0.3$ and $t''=-0.2$ (~\ref{fig:waves-modified}(b)). The rest fo the parameters are as in Fig.~\ref{fig:pares}. Compared with Fig.~\ref{fig:pares}(b), the $\ket{d_{x^2-y^2}}$ state is significantly modified by the additional hopping, resulting in a less localized pair. In contrast, the $\ket{p_x}$ state remains virtually unchanged by $t'$ and $t''$.

%Consider three representative cases. At $J_z = 0.4$, $t = 1$, $t' = t'' = 0$, and $V = 0$, the energy splitting per-hole between $\ket{d_{x^2-y^2}}$ and $\ket{p_x}$ is about $0.043t$. 
%For $t' = 0.03$ and $t'' = -0.07$, this splitting vanishes and the two states become nearly degenerate. Finally, when $t' = -0.1$ and $t'' = 0$, the $\ket{d_{x^2-y^2}}$ state is pushed above the first excited level.

To gain physical insight into the effect of $t'$ and $t''$, we consider the possible hopping processes that allow a hole pair to propagate, as illustrated schematically in Fig.~\ref{fig:propagation}. The hopping $t$ mediates transitions to an excited virtual state, while $t'$ and $t''$ provide alternative connections between intermediate configurations.
Specifically, the $\ket{p_x}$ state can propagate by $(1,0)$ through the intermediate virtual configuration (b1) via two $t$-hopping steps. Note that (b1) is only accessible to $\ket{p_x}$, while its projection onto (c2) vanishes. This corresponds to the sequence
$
(a1)\xrightarrow{\scriptscriptstyle{t}}(b1)\xrightarrow{\scriptscriptstyle{t}}(c1).
$
However, the same final state (c1) can also be reached directly via a single $t''$ hopping.

These two possible paths to (c1) form a three-site loop, which can lead to geometric frustration. The product of the hopping amplitudes around this loop is
$t^2 t''.$
If this product is negative (i.e., $t'' < 0$ for $t > 0$), the energy of a wavefunction occupying this loop is increased. Consequently, we expect a negative $t''$ to inhibit $p$-wave pairing. This view will be supported by our numerical characterization of the hole-pairing symmetries shown below.

In a similar manner, the most probable configuration for $\ket{d_{x^2-y^2}}$ is (b2). This state can propagate by $(1/2,1/2)$ via two alternative paths: either through two $t$-hopping steps,
$
(a1)\xrightarrow{\scriptscriptstyle{t}}(b2)\xrightarrow{\scriptscriptstyle{t}}(c2),
$
or directly through a single $t'$ hopping,
$
(a1)\xrightarrow{\scriptscriptstyle{t'}}(c2).
$
These two paths form a three-site loop with hopping product $t^2 t'$. Since $t > 0$, a negative $t'$ makes this product negative, which increases the energy of a wavefunction occupying the loop---i.e., induces geometric frustration. Therefore, we expect $t' < 0$ to inhibit $d_{x^2-y^2}$-wave pairing, as  predicted by our numerical analysis of the hole-pairing symmetries. 

For the  characterization of the hole-pairing symmetry, we identify the symmetry of the system's states that yield a finite negative pairing energy, and we examine the absence of hole-position correlations. This is done as follows. Let us denote by $\mathcal{E}_o$ the ground-state energy per hole of the interacting two-hole state---for instance, $\mathcal{E}_p$ for the $\ket{p_x}$-like state and $\mathcal{E}_d$ for the $\ket{d_{x^2-y^2}}$-like state---and by $\mathcal{E}_{2h}$ the energy per hole of a completely unpaired reference state, $\ket{2h}$, in which the two holes do not interact. Since the holes are noninteracting in this reference state, its energy per hole is approximately twice the single-hole energy, ${\cal E}_{2h} \approx 2{\cal E}_{1h}$, where ${\cal E}_{1h}$ is the ground-state energy of the Hamiltonian with a single hole. The binding energy per hole is the defined as ${\cal E}_b = {\cal E}_o- {\cal E}_{2h}.$     
A negative value of ${\cal E}_b$ indicates that pairing is energetically favorable relative to the unpaired reference state. We confirm that $\ket{2h}$ is an unpaired state by examining its spatial density correlations; the hole-hole correlation function remains essentially constant across all site separations, $\langle \hat h_{\bm r}\hat h_{\bm 0}\rangle \approx \text{const}$. Consequently, hole pairing is energetically favorable (and thermodynamically stable) whenever ${\cal E}_b < 0$.

%\FM{The hole pairing-symmetry is characterized here by determining the symmetry of the states of the system that yield finite positive pairing energy formation in conjunction with the analysis of the absence of hole correlations with position as follows.   First, we define the binding energy per hole as $e_b = e - e_{2h}$, where $e$ represents the energy per hole of the interacting two-hole states. %(the $\ket{p_x}$ or $\ket{d_{x^2-y^2}}$ states). To evaluate if pairing is energetically favorable, we compare this value against $e_{2h}$, which is the energy per hole of a completely unpaired two-hole state (denoted by $\ket{2h}$). }
%\UD{Because these two holes do not interact in the $\ket{2h}$  state, its energy per-hole is nearly identical to twice the single-hole  ($e_{2h} \approx 2e_{1h}$, where $e_{1h}$ is the ground state energy of the Hamiltonian with only a single hole instead of two). 
%We can confirm that $\ket{2h}$ is an unpaired state by looking at its spatial density correlations: the hole-hole correlation function remains nearly flat and uniform across all positions ($\langle h_{\bm r}h_{\bm 0}\rangle \approx \text{cte}$). Ultimately, hole pairing is thermodynamically stable and favorable whenever $e_b < 0$. 

\begin{figure}[h]
  \includegraphics[page=3]{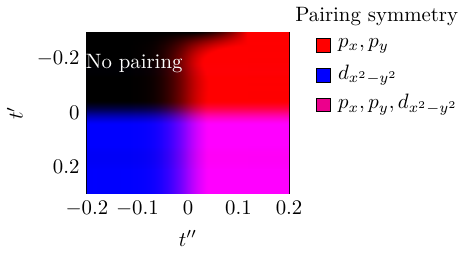}
  \caption{Phase diagram of the pairing symmetry as a function of $t'$ and $t''$ for an $8\times 8$ lattice with two holes ($J=0.4$, $t=1$, and $V=0$). Red and blue regions correspond, respectively, to stable $p$-wave ($\ket{p_x}$ or $\ket{p_y}$) and $d$-wave ($\ket{d_{x^2-y^2}}$) bound pairs with negative binding energy ($\mathcal{E}_b < 0$). Black indicates the absence of hole pairing ($\mathcal{E}_b \ge 0$), while pink marks the region where both pairing symmetries are energetically favored over the unpaired $\ket{2h}$ state.} \label{fig:mapa}
\end{figure}

In Fig.~\ref{fig:mapa}, we present the pairing phase diagram as a function of $t'$ and $t''$ obtained from our numerical calculations for a $8\times 8$ lattice with two holes and $J=0.4,t=1$ and $V=0$. Black regions describes regimes where hole pairing is energetically unfavorable, meaning ${\cal E}_b \ge 0$ across all configurations. Conversely, red (blue) colors mark the parameter space where the $\ket{p_x}$ ($\ket{d_{x^2-y^2}}$) state forms a stable bound pair with a negative binding energy (${\cal E}_b < 0$). The pink region highlights the regime where both the $p$-wave and $d$-wave pairs are simultaneously favorable relative to the unpaired $\ket{2h}$ state. These numerical results are in concordance with our preceding analysis. For instance, for parameters commonly associated with hole-doped cuprates ($t'=-0.3$, $t''=0.2$)\cite{Tohyama_2004,Jiang_2022}, this model predicts a ground state characterized by $p$-wave pairing symmetry and  for electron-doped cuprates $d$-wave symmetry ($t'=0.3$, $t''=-0.2$)\cite{Tohyama_2004,Jiang_2022}.

%%%In Fig.\ref{fig:mapa}, we show the pairing symmetry of the lowest-energy states for different values of $t'$ and $t''$ obtained through numerical calculations. Black represents no pairing in the ground state, and red (blue) colors represent that $\ket{p_x}$  ($\ket{d_{x^2-y^2}}$) symmetry of the pair has lower energy than the first excited unpaired state. Pink color represent when both $p$-wave and $d$-wave pairs have lower energy than the first unpaired state. We can see that our symmetry analysis is correctly reflected in the numerical results. For hole-doped cuprates $t'=0.3$, $t''=-0.2$ within this model, we expect $p$-wave symmetry.

%we show the pairing symmetry of the lowest-energy states for different values of $t'$ and $t''$. Black represents no pairing in the ground state, and red (blue) colors represent that $\ket{p_x}$  ($\ket{d_{x^2-y^2}}$) symmetry of the pair has lower energy than the first unpaired state. We can see that our symmetry analysis is correctly reflected in the numerical results. For hole-doped cuprates ($t'=$ −0.3, $t''=$ 0.2) we should expect p-wave symmetry pairing, and for electron-doped cuprates ($t'=$ 0.3, $t''=$ −0.2) we should expect $d$-wave symmetry. 

%Stripes are electronic patterns found in many two-dimensional (2D) materials.
\subsection{Stripe phase}
As mentioned in the introduction, stripe phases can form in strongly correlated electron materials, most notably in the cuprate high-temperature superconductors. They correspond to a modulated electronic state characterized by coupled unidirectional patterns of charge and spin: charge-density waves (CDW) and spin-density waves (SDW). The wavelengths of these modulations obey the relation \(\lambda_{\text{SDW}} = 2\lambda_{\text{CDW}}\).\cite{Tranquada_1995,Tranquada1997}
%\FMCom{[Add a couple of references]}\UDCom{[In progress]}. 
Such CDW/SDW order separates two distinct antiferromagnetic domains, effectively acting as an antiphase domain wall. In LSCO-based materials, this stripe order has been observed in the doping range $\sim\!0.1 < \delta < 0.135 $ and is strongest at the commensurate doping level $\delta \sim 1/8$ \cite{Simutis_2022}.

%\ vspace{1cm}

\begin{figure}[h]
    \centering
    \includegraphics[page=25]{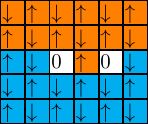}
    \caption{Stripe configuration with linear density $\rho_l =  0.\bar{3}$ on a $5\times 6$ lattice containing two holes. The holes are localized along the antiphase domain wall, thereby reducing the magnetic energy.}
    \label{fig:stripe}
\end{figure}

In what follows, we analyze stripe phases arising within the $t$-$t'$-$t''$-$J_z$ model for systems with a small even number of holes (global hole densities $\delta \lesssim 0.15$) with parameters for a hole-doped cuprate $J=0.4$,  $t=1$, $t'=-0.3$ and $t''=0.2$.  Specifically, we examine $5 \times M$ lattices, with $M$ even and $M > 5$, under periodic boundary conditions. Two sets of calculations are performed: (i) using 4 holes, which yields linear hole densities along the stripe $\rho_l = 4/M = 0.4, 0.5, 2/3$ for $M = 10, 8, 6$, respectively; and (ii) using 2 holes, giving $\rho_l = 2/M = 1/3, 0.25, 0.2$ for $M = 6, 8, 10$, respectively.  
%\UDCom{The remaining parameters $t'$, $t''$, and $J_z$ are set as in Fig.~\ref{fig:parameters}} \UDCom{[We may need to remark that these parameters are for a typical hole-doped cuprate as LSCO.]}. 
Our calculations show that such geometry configuration and set of parameters yields the formation of a horizontal antiphase domain wall of length $M$, characterized by a line of \textit{wrong bonds} where  neighboring sites along the vertical direction have the same spin orientation, {\it i.e.},  $S^z(x,y)S^z(x,y+1)=1/4$ for $x\in\{0,1,\ldots,L-1\}$ for some $0\le y<M$, instead of the staggered spin configuration. As an example we show in Fig.\ref{fig:stripe} the ground state spin configuration for a $5\times6$ lattice with two holes ($\delta=0.067$) thats shows a linear stripe with $\rho_l=1/3$.
  
\vspace{0.2cm}

%%In this work, we \FM{will focus in the study of}\st{only analyze} stripes with an even number of holes. 

\begin{comment}

In Fig. \ref{fig:stripes} we show the \FM{calculated} most probable configuration for the striped ground state of a system with four holes in an $8\times4$ lattice \FM{under the }parameters \FM{simulation $t=1,t'=-0.3,t''=0.2$ and $J=0.4$}. \FM{The two possible antiferromagnetic domains, colored blue and orange, correspond to regions whose spins are mutually rotated by $180^{\circ}$.} The holes are aligned along the two antiphase domain walls oriented along the $y$-axis, \FM{with two holes sitting at each wall.}%(two holes per wall).
\UD{}
\begin{figure}[h]
    \centering   
\includegraphics[page=3,width=0.20\textwidth]{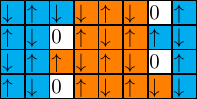} 
    \caption{Most probable configuration for a striped ground state (four holes in an $8\times4$ lattice). Holes separate the two antiferromagnetic domains (orange and blue).}
    \label{fig:stripes}
\end{figure}
\end{comment}

Regarding the energy cost of adding holes to a unidirectional stripe, Chernyshev et al.~\cite{Chernyshev_2002} analyzed this problem within the $t$-$J_z$ model under the rigid band approximation. They found that the process is equivalent to adding holes to a strictly one-dimensional system, and the resulting energy is given by
\begin{align}
{\cal E}_S = \frac{J_z}{2\rho_l} + A\cos(\pi\rho_l) + B,
\label{eq:energystripe}
\end{align}
where $A$ and $B$ are constants for fixed $J_z$ and $t$. In this expression, the first term corresponds to the domain-wall energy, while the second term arises from the kinetic energy of free particles filling the one-dimensional band up to the Fermi wavevector $k_F = \pi\rho_l$~\cite{Chernyshev_2002}.

\begin{figure}[h]
  \includegraphics[page=33]{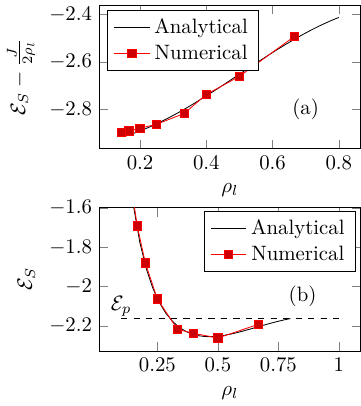}
  \vspace{-0.15cm}
  \caption{Comparison of the energy cost to add holes to a unidirectional stripe as a function of the linear hole density $\rho_l$, contrasting the rigid-band approximation  with numerical estimates. (a) Within the rigid approximation, hole addition to the antiphase domain wall follows a simple one-dimensional filling rule. (b) The numerically computed energy per hole in a single stripe  $\mathcal{E}_S$ (for $J=0.4$, $t=1$, $t'=-0.3$, and $t''=0.2$; remaining parameters are given in the main text) exhibits a minimum at $\rho_l \approx 0.5$. At this density, $\mathcal{E}_S$ is approximately $0.1 t$ lower than the pair energy, i.e., $\mathcal{E}_S - \mathcal{E}_p \approx -0.1 t$.
  %\caption{(a) The addition of holes to the antiphase domain wall follows a simple rule similar to a one-dimensional system. (b) The energy per hole ${\cal E}_S$ for parameters $J=0.4,t=1,t'=-0.3,t''=0.2$. ${\cal E}_S$ has a minimum at $\rho_l\approx 0.5$ and is about $0.1t$ lower than that of a pair (${\cal E}_S-{\cal E}_p\approx -0.1t$).\FMCom{[Which values of M you use here?]}
  }
  \label{fig:stripe-energy}
\end{figure}

We have calculated the energy per hole ${\cal E}_S$  for a single stripe, as well as the one shifted respect to the domain wall energy, (${\cal E}_S-J_z/2\rho_l$) as a function of  hole filling $\rho_l$ in $5\times M$ lattices ($M$ even). A comparison of our numerical calculations with the analytical results of Eq.~(\ref{eq:energystripe}) gives us a reasonable good agreement by fitting $A=-0.296$  and $B=-2.652$ (Fig.~\ref{fig:stripe-energy})
It is notable the agreement despite the fact that including  $t'$ and $t''$ hoppings goes beyond the rigid band approximation.

The optimal filling that minimizes $\mathcal{E}_S$ for a stripe is $\rho_l^{\text{op}} \approx 0.5$, consistent with other numerical results (Fig.~\ref{fig:stripe-energy}(b)) \cite{White_1998_2,Corboz_2011,Marino_2022}. A dashed horizontal line is drawn at 
the ground-state energy (per
hole) of the interacting pair state in the dilute regime, $\mathcal{E}_p = -2.16$. The energy difference (per hole) between the optimally filled stripe and the pair is therefore $\mathcal{E}_S - \mathcal{E}_p \approx -0.1$. 
%\UD{At $T=0$,} holes tend to form the lowest-energy structures . \FMCom{[So what???. Not sure what do you want to express here. I do not understand. A more clear explanation is needed of what is the physical consequence that $\mathcal{E}_S - \mathcal{E}_p < 0$.  ] }

%In large systems, and in the absence of long-range interactions, they will form half-filled stripes with separation $L \ge 5$ whenever possible. Thus, the half-filled ($\rho_l = 0.5$) striped phase is the ground state for $\delta = \rho_l/L \le 0.1$. 
%\UD{At $T=0$,} \FMCom{All your calculations are at $T=0$. Not need to specify.} 
The numerical results show that holes will tend to form the structures of lowest possible energy, and in large systems, in the absence for long-range interactions, they will try to form half-filled stripes with separation $L\ge 5$, whenever possible. Thus, the half-filled $\rho_l=0.5$ striped phase is the ground state for $\delta=\rho_l/L\le 0.1$. 
Similarly, a stripe with $0.5 \le \rho_l < 0.75$ has lower energy than hole pairs (or unpaired holes), so the ground state remains striped with $\rho_l \ge 0.5$ for doping $0.1 \le \delta \le 0.15$.

\vspace{0.2cm}

\subsection{Long-range repulsive interaction}

%We have found that the stripe near half-filling is always the ground state within this model in the absence of electron repulsion ($V=0$). To change this, we must introduce additional interactions. 
We now extend the $t$-$t'$-$t''$-$J_z$ model by including electron-electron interactions beyond the on-site Hubbard $U$, namely long-range Coulomb or short-range Yukawa/dipolar potentials. Let us first analyze the exchange energy of a system containing two holes in the absence of such interactions, in a simplified but illustrative manner. For two holes at adjacent sites ($r = 1$), the number of broken antiferromagnetic bonds relative to the N\'eel state is seven, corresponding to a magnetic energy cost of $7J/2$. For $r > 1$, the minimal magnetic cost is $8J/2$. Thus, bringing the holes close together reduces the magnetic energy by $J/2$. Adding a repulsive electric potential modifies this energy landscape. The total potential energy can be written then as the sum of the magnetic and electric contributions, $E_{\text{pot}} = E_M + E_V$, where $E_M = \bra{\Psi} H_{J_z} \ket{\Psi}$, and 
\begin{align}
 E_V = \frac{V}{2} \bra{\Psi} \sum_{\bm j \neq \bm i} f(|\bm i - \bm j|) \, \hat n_{\bm i} \hat n_{\bm j} \ket{\Psi},   
\end{align}
\noindent with $f(r)$ a function of the hole separation $r = |\bm i - \bm j|$.

\begin{figure}[h!]
  \includegraphics[page=37]{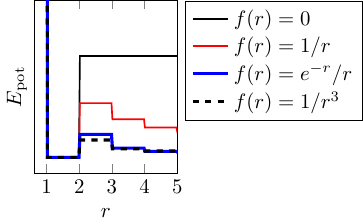}
  \caption{Total potential energy (in a.u.) as a function of hole separation $r$ (in lattice units), computed for various repulsive interaction potentials with $V = J/2$. The minima of the potential energy pockets have been aligned for direct comparison.
  %Schematic representation of the potential energy (magnetic plus electric) for different repulsive interactions with $V=J/2$. The floors of the \textit{pockets} have been leveled for comparison.
  }
  \label{fig:schematic}
\end{figure}

We consider several functional forms for the repulsive potential: $f(r) = 0$ (no repulsion), $f(r) = 1/r$ (long-range Coulomb), $f(r) = e^{-r}/r$ (Yukawa, corresponding to Thomas--Fermi screening), and $f(r) = 1/r^3$ (dipolar-like interaction).
For $f(r) = 0$, the total potential energy $E_{\text{pot}}(r)$ exhibits a characteristic \textit{pocket} with a minimum at $r = 1$, giving rise to an effective attractive force between the holes (Fig.\ref{fig:schematic}). This attraction can be strong enough to drive phase separation. Introducing a short-range repulsive potential (such as Yukawa or dipolar) tends to fill or eliminate this pocket, depending on the value of $V$, thereby suppressing pairing. By contrast, the long-range Coulomb repulsion, which decays more slowly, appears to be the most suitable candidate for modifying the energy landscape while preserving a viable pairing mechanism.

%Figure \ref{fig:schematic} shows a simplified diagram of the total potential energy versus hole separation for different functions: $f(r)=0$ (no repulsion), $f(r)=1/r$ (Coulomb long-range), $f(r)=e^{-r}/r$ (Yukawa, corresponding to Thomas-Fermi screening), and $f(r)=1/r^3$ (dipolar-like screening). Note that for $f(r)=0$, the potential forms a \textit{pocket} with an attractive force at $r=1$. This attraction can be strong and cause phase separation. On the other hand, short-range repulsive potentials can lead to a very shallow pocket or no pocket at all (depending on $V$), inhibiting pairing. The long-range Coulomb repulsion seems more suitable in this case. 

We have found that for two holes in a $6\times6$ lattice ($\delta=0.055$) with $J=0.4,t=1,t'=-0.3,t''=0.2$, the ground state is paired up to $V=2.7$. This result is close to the estimate by Calzado et al.\cite{Calzado_2001}. For two holes in an $8\times8$ lattice ($\delta=0.031$), the ground state is paired up to $V\approx 1$. Given that it is known that superconductivity appears in LSCO materials at $\delta=0.055$, we consider $V\approx 3$ to be a realistic value, with $V=1$ as a lower bound.

\begin{figure}[h!]
\includegraphics[page=24,width=0.30\textwidth]{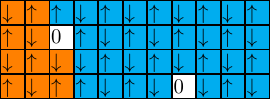}
  \caption{Ground-state configuration showing holes expelled from the half-filled stripe upon the inclusion of the long-range repulsive interaction at $V \approx 1.8$, for parameters $J = 0.4$, $t = 1$, $t' = -0.3$, and $t'' = 0.2$ on an $11\times4$ lattice.
  %Holes are expelled from the half-filled stripe upon inclusion of the long-range repulsive interaction at $V\approx1.8$, $J=0.4$, $t=1$, $t'=-0.3$, $t''=0.2$ in an $11\times4$ lattice.}\label{fig:hole-expulsion
  }
  \label{fig:hole-expulsion}
\end{figure}

For the stripe geometry, specifically on an $11\times4$ lattice with line density $\rho_l = 0.5$, we find that the long-range Coulomb interaction expels a hole from the stripe already at $V \approx 1.8$ (see Fig.~\ref{fig:hole-expulsion}). In contrast, short-range interactions, such as $1/r^3$ and Yukawa potentials, require a higher coupling, $V \approx 2.3$, to induce hole expulsion. For purely next-nearest-neighbor repulsion, the threshold increases even further to $V > 6$. Thus, long-range interactions destabilize the half-filled stripe, whereas shorter-range interactions tend to stabilize it. Moreover, since next-nearest-neighbor repulsion alone cannot expel a hole for $\rho_l \le 0.5$ at realistic values of $V$---and short-range repulsions generally have a weaker effect at lower densities---the stripe phase is expected to dominate at low doping ($\delta < 0.10$) in the extended Hubbard model.

%In the case of a stripe, we found that for two holes in an $11\times4$ lattice ($\rho_l=0.5$), the long-range repulsive interaction causes the expulsion of a hole from the stripe at $V\approx1.8$ (see Fig.~\ref{fig:hole-expulsion}). On the other hand, short-range interactions (e.g., $1/r^3$ and Yukawa) require $V\approx 2.3$ to expel the hole, and for next-nearest-neighbor repulsion only, we even need $V>6$. Thus, the long-range interaction makes the half-filled stripe unstable, whereas for shorter-range interactions, the stripe phase is more stable. Since the next-nearest-neighbor interaction alone cannot expel a hole for $\rho_l\le0.5$ for reasonable values of $V$ (short-range repulsive interactions have a weaker effect on stripes at lower densities), the stripe phase will dominate at low doping ($\delta<0.10$) in the extended-Hubbard model.

%interactions do not, even for reasonable values of $V$.

\subsection{Pair expulsion from the stripe}

\begin{figure}[h]
  \subfigure[$\ket{S}$ pure stripe]{\includegraphics[page=15]{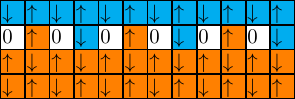}}
  \subfigure[$\ket{S+p}$ stripe plus pair] {\includegraphics[page=14]{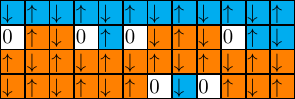}}
  \caption{Two characteristic ground-state configurations are shown: (a) a pure stripe, $\ket{S}$, and (b) a stripe containing an expelled hole pair, $\ket{S+p}$, oriented parallel to the stripe.
  %There are two proposals for the ground state: (a) a pure stripe $\ket{S}$, and (b) a stripe with an expelled pair $\ket{S+p}$. The pair has the same orientation as the stripe.
  }
  \label{fig:stripepluspair}
\end{figure}

Let us consider two characteristic ground states. One is the pure stripe state, which is the ground state near half-filling at $V=0$, and denoted by $\ket{S}$ with energy per hole $\mathcal{E}_S$ . The other, denoted $\ket{S+p}$, consists of a stripe plus an expelled hole pair oriented parallel to it with ground state $\mathcal{E}_{S+p}$ (See Fig.\ref{fig:stripepluspair}).
Using these two states as trial wavefunctions, we apply then the Lanczos-based algorithm~\cite{Chiappa_2001} for increasing values of the truncation parameter $n$, up to $n\sim 10^7$. We consider lattice sizes $12\times7$, $12\times5$, $10\times5$, and $8\times5$, each containing six holes, which correspond to dopings $\delta = 0.07, 0.10, 0.12,$ and $0.15$, respectively.

%Let us consider two characteristic ground states; one a pure stripe state is the ground state near half-filling when $V=0$ and is denoted $\ket{S}$, with energy per hole ${\cal E}_S$. The other state, denoted $\ket{S+p}$, has an expelled pair oriented similarly to the stripe. 

%We then applied the Lanczos-based algorithm~\cite{Chiappa_2001} for increasing numbers of states $n$ for the two trial wavefunctions, starting from configurations similar to those described above (Fig.~\ref{fig:stripepluspair}) for lattice sizes $12\times7$, $12\times5$, $10\times5$, and $8\times5$ with six holes ($\delta=0.07,0.10,0.12,0.15$, respectively).

%Let us consider two characteristic pure stripe state is the ground state near half-filling when $V=0$ and is denoted $\ket{S}$, with energy per hole ${\cal E}_S$. The other state, denoted $\ket{S+p}$, has an expelled pair oriented similarly to the stripe. We applied a Lanczos-based algorithm~\cite{Chiappa_2001} for increasing numbers of states $n$ for the two trial wavefunctions, starting from configurations similar to those in Fig.~\ref{fig:stripepluspair} for lattice sizes $12\times7$, $12\times5$, $10\times5$, and $8\times5$ with six holes ($\delta=0.07,0.10,0.12,0.15$, respectively).

%We compare the energies of two states for different lattice sizes: a pure stripe and a stripe with an expelled pair (see Fig.~\ref{fig:stripepluspair}).

\begin{figure}[h]
  \includegraphics[page=18,width=0.49\textwidth]{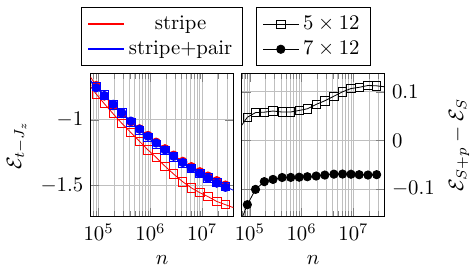}
  \includegraphics[page=19,width=0.49\textwidth]{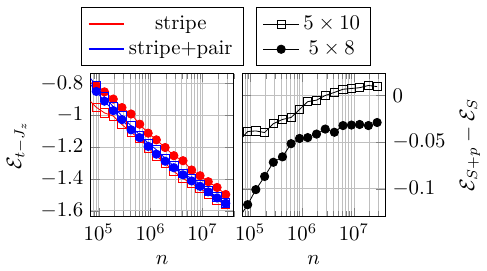}
\caption{Energy per hole, $\mathcal{E}_{t-J_z} = \mathcal{E} - \mathcal{E}_V$, and energy difference $\Delta \mathcal{E} = \mathcal{E}_{S+p} - \mathcal{E}_S$ between the pure stripe state $\ket{S}$ and the stripe-plus-pair state $\ket{S+p}$. (a) Results for dopings $\delta = 0.07$ ($12\times7$ lattice) and $\delta = 0.10$ ($12\times5$ lattice). (b) Results for $\delta = 0.12$ ($10\times5$ lattice) and $\delta = 0.15$ ($8\times5$ lattice). All systems contain six holes.
%Energy per hole \FM{${\cal E}_{t-J_z}={\cal E}-{\cal E}_V$} and energy difference ${\cal E}_{S+p}-{\cal E}_S$ for the stripe $\ket{S}$ and stripe-plus-pair $\ket{S+p}$ states for (a) $\delta=0.07$ ($12\times7$ lattice) and $\delta=0.10$ ($12\times5$ lattice), and (b) $\delta=0.12$ ($10\times5$ lattice) and $\delta=0.15$ ($8\times5$ lattice). All systems have six holes.}\label{fig:convergence
}
\label{fig:convergence}
\end{figure}

We find that the energy of a trial state follows a power law in the number of retained states, 
$$
\mathcal{E}(n) = \mathcal{E}^\infty + b n^C,
$$
where $\mathcal{E}^\infty$ is the true ground-state energy, and $b$ and $C$ are constants. Consequently, the energy difference between the two competing states as a function of $n$ is
$$
\mathcal{E}_{S+p}(n) - \mathcal{E}_{S}(n) = \mathcal{E}_{S+p}^\infty - \mathcal{E}_S^\infty + b_1 n^{C_1} - b_2 n^{C_2}.
$$
For sufficiently large $n$, the term with the largest exponent dominates. In Fig.~\ref{fig:convergence}, we show the evolution of the variational energy ${\cal E}_{t-J_z}$ and the energy difference $\mathcal{E}_S - \mathcal{E}_{S+p}$ for the two competing states with six holes and parameters $V = 3$, $J = 0.4$, $t = 1$, $t' = -0.3$, and $t'' = 0.2$.
For $\delta = 0.10$ ($12\times5$ lattice, stripe filling $\rho_l = 0.5$), the pure stripe state has lower energy than the stripe-plus-pair state. In this case, the reduction in potential energy from expelling the pair is insufficient to overcome the energy cost of disrupting the stripe. Conversely, for $\delta = 0.07$ ($12\times7$ lattice, also with $\rho_l = 0.5$), the increased system width of 7 makes hole expulsion energetically favorable. At $\delta = 0.12$ ($10\times5$ lattice), the two states become nearly degenerate, whereas for $\delta = 0.15$ ($8\times5$ lattice), the stripe-plus-pair state is actually lower in energy than the pure stripe. For these latter two dopings, the stripe filling is higher than the optimal $\rho_l = 0.5$; expelling holes reduces the filling within the stripe, which stabilizes the remaining holes inside and allows the expelled pairs to exist outside the stripe.

This charge expulsion from the stripe, enabled by the long-range repulsive interaction, constitutes the key mechanism for superconductivity in this model over the doping range $0.05 < \delta < 0.15$, as we will show below through the pairing correlation calculations.

%%\UDCom{[In ${\cal E}^{t-J_z}$, the superscript is not an error. It is to avoid collision with the subscript used to indicate the type of state, ${\cal E}^V$ is the repulsive energy, ${\cal E}^V_S$ is the repulsive energy of an striped state.]} 

%For $\delta=0.10$ ($12\times5$ lattice, stripe filling $\rho_l=0.5$), the stripe has lower energy than the stripe plus a pair. This is because the reduction in potential energy from expelling the pair is not sufficient to favor the stripe-plus-pair state. In contrast, for $\delta=0.07$ (stripe filling also $\rho_l=0.5$), the system width of 7 is sufficient to make hole expulsion favorable. For $\delta=0.12$, both states have nearly the same energy, and for $\delta=0.15$, the stripe-plus-pair state is lower in energy than the pure stripe. In these cases, the stripe filling is higher than optimal, so expelling holes from the stripe benefits the remaining holes inside the stripe, stabilizing the pairs outside.

%The charge expulsion from the stripe, enabled by long-range repulsive interactions, is the key ingredient for understanding superconductivity in this model for doping values $0.05<\delta<0.15$ as shown below through the pairing correlation calculations. 

\subsection{Pairing correlations}
We have already shown that the holes expelled from the stripe by the long-range repulsion form bound states ($\mathcal{E}_b < 0$). The remaining question is whether these bound states coherently propagate and establish long-range phase coherence---{\it i.e.}, whether they exhibit superconducting behavior.

To probe superconductivity in these states, we have numerically calculated the triplet pair--pair correlation function\cite{Moreo_1991},
\begin{align}
P_{xx}(\bm r) = \langle \Delta_x^\dagger(\bm r) \Delta_x(0) \rangle,
\end{align}
where
\begin{align}
\Delta_x(\bm r) = \frac{1}{\sqrt{2}} \left( c_{\bm r+\hat{e}_x,\uparrow} c_{\bm r,\downarrow} + c_{\bm r+\hat{e}_x,\downarrow} c_{\bm r,\uparrow} \right).
\end{align}
%\FMCom{[Before you had a minus sign there in (20). I expect that in the actual numerical calculation you use this definition and not the one you have before]}\UDCom{[Yes, I actually used the definition and notation by Moreo et. al, but apparently nobody uses it now but they are equivalent.]}
The operator $\Delta_x^\dagger(\bm r)$ creates a spin-triplet Cooper pair on the nearest-neighbor bond along the $x$-direction, centered at site $\bm r$. The function $P_{xx}(\bm r)$ thus measures the spatial coherence of pairing, {\it i.e.}, the correlation between the amplitude to create a pair at the origin and the amplitude to annihilate a pair at position $\bm r$.

The pair--pair correlation function for a given distance $|\bm r|$ is then defined as
\begin{align}
P_{xx}(|\bm r|) = \sum_{|\bm r'| = |\bm r|} \frac{P_{xx}(\bm r')}{N_{|\bm r|}},
\end{align}
where $N_{|\bm r|}$ is the number of lattice vectors with length $|\bm r|$. In a superconducting state, $P_{xx}(|\bm r|)$ should decay at most as a power law as $|\bm r| \to \infty$, whereas in a non-superconducting state---such as an insulator or a normal metal with local pairing---it decays exponentially \cite{Yang_1962, Kosterlitz_1973}.

%For a translationally invariant system, $P_{xx}(\bm r)$ corresponds to $p_x$ symmetry\cite{Moreo_1991}. 
%(it annihilates a $p_x$ pair at the origin and creates one at $\bm r$, see appendix~\ref{sec:appendix-pair} for the proof). 
%the sum of $P_{xx}(\bm r')$ over all vectors $\bm r'$ with the same distance $|\bm r|$, divided by the number of such vectors $N_{|\bm r|}$:

% We have plotted $P_{xx}(|\bm r|)$ in Fig.~\ref{fig:correlation} for the horizontally aligned pair $\ket{p_x}$, the vertically aligned pair $\ket{p_y}$, two unpaired holes $\ket{2h}$, and the four-hole half-filled stripe $\ket{S}$ in an $8\times5$ lattice. As can be seen, $P_{xx}(|\bm r|)$ decays exponentially for all states except the pair $\ket{p_x}$, which exhibits off-diagonal long-range order (ODLRO) and thus superconductivity. This true ODLRO is an effect of having discarded the spin dynamics. A similar plot for $P_{yy}(|\bm r|)$ shows that $\ket{p_y}$ also displays superconductivity. We found that stripes with $\rho_l=1$ and $\rho_l\le0.5$ show no signs of superconductivity, but surprisingly, stripes with $\rho_l=0.\bar{6}$ do. 

\begin{figure}[h]
 \includegraphics[page=10]{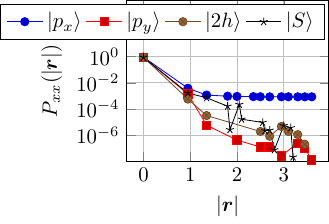}
  \caption{The pair-pair correlation function $P_{xx}(r)$ for four states: the pairs $\ket{p_x}$ and $\ket{p_y}$, two unpaired holes $\ket{2h}$, and a half-filled stripe $\ket{S}$. Note that $P_{xx}(|\bm r|)$ decays exponentially for all states except for the $\ket{p_x}$ pair (blue dots), which exhibits off-diagonal long-range order, signature of superconductivity behavior.}\label{fig:correlation}
\end{figure}

\begin{figure}[h]
 \subfigure[]{\includegraphics[page=30]{figuras-superconductividad.pdf}}
 \subfigure[]{%%\includegraphics[page=21]{figuras-superconductividad.pdf}
 \includegraphics[]{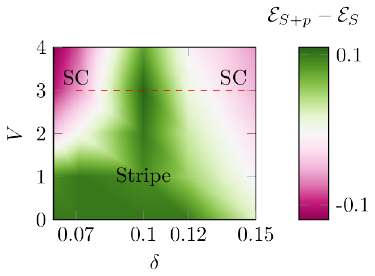}
 
 }
 \caption{Phase diagram showing: (a) Pair-pair correlation $P_{xx}(|\bm r|)$. The horizontal stripe $\ket{S}$ with $\rho_l = 0.\bar{6}$ shows a non-zero $P_{xx}$ asymptote, indicating $p_x$-pair exchange and superconducting character similar to $\ket{S+p_x}$; the stripe with a vertical pair $\ket{S+p_y}$ instead decays exponentially (no $p_x$ pairing). (b) Superconducting phase diagram: doping $\delta$ vs repulsion $V$ for $J=0.4$, $t=1$, $t'=-0.3$, $t''=0.2$. At $V=3$, the system shows reentrant behavior reminiscent of Nd-LSCO (red dashed-line).
 }
 \label{fig:phase-diagram}
\end{figure}

We plot $P_{xx}(|\bm r|)$ in Fig.~\ref{fig:correlation} for the horizontally aligned pair $\ket{p_x}$, the vertically aligned pair $\ket{p_y}$, two unpaired holes $\ket{2h}$, and the half-filled four-hole stripe $\ket{S}$, all on an $8\times5$ lattice. As shown in the figure, $P_{xx}(|\bm r|)$ decays exponentially for all states except for the $\ket{p_x}$ pair, which exhibits off-diagonal long-range order (ODLRO)---a hallmark of superconductivity. We attribute this true ODLRO to the neglect of spin dynamics in the model. A similar analysis of $P_{yy}(|\bm r|)$ reveals that $\ket{p_y}$ also displays superconducting behavior. Interestingly, while stripes with $\rho_l = 1$ and $\rho_l \le 0.5$ show no signs of superconductivity, we find that stripes with filling $\rho_l = 0.\bar{6}$ do exhibit superconducting features.

In Fig.~\ref{fig:phase-diagram}(a), we show $P_{xx}(|\bm r|)$ for three states: a horizontal stripe with filling $\rho_l = 0.\bar{6}$ ($\ket{S}$), a horizontal stripe plus a horizontally aligned pair ($\ket{S+p_x}$), and a horizontal stripe with a vertically aligned pair ($\ket{S+p_y}$). The correlation function for $\ket{S+p_x}$ indicates that $p_x$ pairs can coexist on opposite sides of the system, as $P_{xx}(|\bm r|)$ does not vanish as $|\bm r| \to \infty$. Remarkably, the pure stripe $\ket{S}$ with $\rho_l = 0.\bar{6}$ also exhibits a non-zero asymptotic value of $P_{xx}(|\bm r|)$, suggesting that it exchanges $p_x$ pairs with the surrounding medium---since the pairs and the stripe holes occupy opposite sides of the lattice. In fact, the superconducting properties of the pure stripe $\ket{S}$ are comparable to those of the stripe with a $p_x$ pair, $\ket{S+p_x}$. In contrast, the stripe with a vertically aligned pair, $\ket{S+p_y}$, shows an exponentially decaying $P_{xx}(|\bm r|)$, indicating no $p_x$-pairing character.

Finally, in Fig.~\ref{fig:phase-diagram}(b), we present the superconducting phase diagram as a function of doping $\delta$ and repulsive interaction strength $V$, for fixed parameters $J = 0.4$, $t = 1$, $t' = -0.3$, and $t'' = 0.2$. The phase diagram is constructed by plotting the energy difference $\Delta E = \mathcal{E}_{S+p} - \mathcal{E}_S$ between the pure stripe state $\ket{S}$ and the stripe-plus-pair state $\ket{S+p}$. A negative value of $\Delta E$ indicates that the stripe-plus-pair state (i.e., a superconducting-like phase) is energetically favored.

At $V = 0$, the pure stripe phase is dominant across all dopings. In contrast, for $V = 3$, we observe a reentrant behavior: the system is superconducting-like at $\delta \approx 0.07$, becomes non-superconducting (half-filled stripe dominant) in the doping range $0.10 \lesssim \delta \lesssim 0.12$, and re-enters the superconducting regime for $\delta > 0.12$. This sequence of phases closely resembles the behavior observed in Nd-LSCO (neodymium-doped La$_{2-x}$Sr$_x$CuO$_4$) \cite{Chen_2025}.

\section{Concluding Remarks}

In this work, we have investigated the pairing symmetry and superconducting properties of an extended two-dimensional $t$-$t'$-$t''$-$J_z$ model, treating anisotropic Ising-like magnetic interactions and long-range electron--electron repulsions within an unified framework. We show that depending upon the sign and magnitude of $t'$ and $t''$ hopping parameters a dominant  $p$-wave, $d$-wave or coexistence of both symmetries may be present, underscoring the crucial role of $t'$ and $t''$ in determining the pairing channel. We further find that long-range repulsive interactions in the low doping regime ($\delta\approx0.7$) can expel hole pairs from these non-superconducting stripe domains, thereby stabilizing a superconducting phase.

%Symmetry-based analysis further reveals that the pair symmetry is $p$-wave for hole-doped cuprates and $d_{x^2-y^2}$-wave for the electron-doped counterpart, underscoring the crucial role of $t'$ and $t''$ in determining the pairing channel.

The calculated superconducting phase diagram exhibits a behavior that closely resembles experimental observations in hole-doped cuprates, such as Nd-LSCO (neodymium-doped La$_{2-x}$Sr$_x$CuO$_4$) \cite{Chen_2025}. Moreover, our pairing correlation calculations strongly indicate that repulsive Coulomb interactions drive reentrant superconductivity at doping levels between $\delta \approx 0.07$ and $\delta \approx 0.15$, consistent with phenomena reported in niobium-based cuprates. We believe that these results highlight the essential role of long-range electrostatic repulsions in strongly correlated systems and provide a solid groundwork for further theoretical analyses of unconventional superconducting mechanisms.

%We have shown that long-range repulsive interactions can be a key ingredient for understanding superconductivity in hole-doped cuprates. In the proposed model with second- and third-neighbor hoppings, Ising-like magnetic interactions, and long-range repulsion, pairs of holes can be expelled from non-superconducting stripes, giving rise to superconductivity. From a symmetry analysis, we found that the pair symmetry is $p$-wave for hole-doped cuprates and $d$-wave for electron-doped cuprates. The calculated superconducting phase diagram resembles that experimentally observed in two-dimensional hole-doped cuprates such as Nd-LSCO. We believe this work highlights the importance of long-range repulsive interactions and can serve as a starting point for analyzing other models.

\section{Acknowledgments}
%F. Mireles \UD{and U.A.D.R} acknowledge funding from PAPIIT-DGAPA-UNAM project IN113920. U. A. Diaz-Reynoso acknowledges the UNAM-DGAPA postdoctoral scholarship. \FMCom{[Tu viaje al APS MM2026 fue pagado por este proyecto también]}\UD{[Y si, muchas gracias por su apoyo.]}\FMCom{[{\bf jajajaja ...de nada, pero no te pedía que me agradecieras, sino que agradecieras explícitamente al proyecto por apoyarte. Lee abajo EN AZUL lo que esperaba escribieras.} ]}This work was supported by the Universidad Nacional Autónoma de México Postdoctoral Program (POSDOC).
%\vspace{1cm}

F. M. and U. A. D.-R. acknowledge funding from PAPIIT-DGAPA-UNAM project IN111624. U. A. D.-R. further acknowledges support from the UNAM-DGAPA postdoctoral scholarship.

\appendix 

\section*{Appendix A: \\ Symmetry operations and pair counting}

We consider an $M\times M$ square lattice ($M$ even, $M\ge 4$) with periodic boundary conditions. The point group $D_4$ is generated by the diagonal reflection $s:(x,y)\to(y,x)$ and the 90$^\circ$ rotation $r:(x,y)\to(y,-x)$. In the main text we use the operations
\begin{align} s: (x,y) &\to (y,x) \nonumber \\ r^2: (x,y) &\to (-x,-y) \nonumber \\ sr: (x,y) &\to (-x,y) \nonumber
\end{align}

The staggered antiferromagnetic background is defined by $S^z(x,y)=(-1)^{x+y}/2$, and is invariant under all three operations, since $-x$ and $-y$ have the same parity as $x$ and $y$.

The three hole configurations of Fig.~1 are invariant under the following operations, yielding the corresponding numbers of exchanged electron pairs $p$:

\medskip
\noindent
{\it Configuration} (a) $\ket{\phi_{(a)}}$: holes at $(0,1)$ and $(0,-1)$; $S^z(0,0)=-1/2$. Invariant under $r^2$. Fixed sites: $(0,0)$, $(M/2,0)$, $(0,M/2)$, $(M/2,M/2)$. Total electrons $M^2-2$, exchanged electrons $(M^2-2)-4=M^2-6$, hence
\[
p_{r^2}=\frac{M^2-6}{2}.
\]

\medskip
\noindent
{\it Configuration} (b) $\ket{\phi_{(b)}}$: holes at $(0,0)$ and $(0,1)$. Invariant under $sr$. Fixed columns: $x=0$ and $x=M/2$, containing $2M$ sites; two holes lie on these columns, so invariant electrons are $2M-2$. Exchanged electrons $(M^2-2)-(2M-2)=M(M-2)$, hence
\[
p_{sr}=\frac{M(M-2)}{2}.
\]

\medskip
\noindent
{\it Configuration} (c) $\ket{\phi_{(c)}}$: holes at $(1,0)$ and $(0,1)$; $S^z(0,0)=-1/2$. Invariant under $s$. Fixed diagonal: $M$ sites. Off-diagonal electrons $(M^2-M)-2=M^2-M-2$, hence
\[
p_{s}=\frac{M^2-M-2}{2}.
\]


\begin{thebibliography}{99}
\bibitem{Footnote1} Stripe ordering correspond to unidirectional modulations of charge and spin density that emerge in the underdoped regime, particularly near $\delta=1/8$, where holes segregate into one-dimensional domain walls separated by antiferromagnetically ordered regions.
\end{thebibliography}
\end{document}